\documentclass[aps,prd,eprint,groupedaddress,showkeys,showpacs]{revtex4-1}

\usepackage{graphicx}
\usepackage{amsmath}
\usepackage{txfonts}
\usepackage{balance}
\usepackage{textcase}
\usepackage{float}
\usepackage{subcaption}

\begin{document}

\title{Frame-dragging effects in obliquely rotating magnetars}

\date{\today}

\author{Debojoti Kuzur} 
\affiliation{Indian Institute of Science Education and Research Bhopal, Bhopal, India}
\author{Ritam Mallick}
\email{mallick@iiserb.ac.in}
\affiliation{Indian Institute of Science Education and Research Bhopal, Bhopal, India}

\begin{abstract}
Magnetars are highly magnetized neutron stars. For a slowly rotating magnetar, the strong magnetic field deforms the star, making it axisymmetric with respect to the magnetic axis (the body symmetry axis). In magnetars, the rotation axis is tilted to the magnetic axis, and we have an oblique rotator. General relativistic treatment of the obliquely rotating magnetar gives rise to frame-dragging velocities both in the azimuthal and polar direction. Solving the Einstein equation up to first-order perturbation in rotation and second-order perturbation in the magnetic field, we calculate the geodesic of a particle near the star's surface. The polar frame-dragging velocity makes the particle orbit non-planar, and the particle moves both along the azimuthal and polar direction for a fixed radial distance. The extent of particle deviation from planar orbit depends on the magnetic field strength and the misalignment angle. We find that the continuous gravitational wave emitted from such obliquely rotating axisymmetric star is non zero, and for small misalignment angle, the gravitational wave amplitude depends more on the azimuthal frame-dragging velocity. In contrast, for a large misalignment angle, the polar frame-dragging velocity dominates. The energy loss from such a misaligned rotator depends more significantly on the polar frame-dragging velocity and therefore, can significantly affect the magnetosphere around a magnetar.
\end{abstract}



\maketitle

\section{Introduction}
The discovery of gravitational wave (GW) GW170817 \cite{abbott} has opened the door of multi-messenger astronomy, especially in connection with neutron stars (NSs). The theory of NS was initially proposed in 1934 \cite{bade}. Soon after this, NS were observationally connected with pulsars \cite{hewish}. Pulsars are rotating NS, emitting out electromagnetic rays (mostly in the radio and x-ray bands) from their magnetic poles. They were first observed as sharp, intense, fast, and very regular pulses of radio waves coming from great distances. The pulse period ranges from a few milliseconds to seconds. The origin of the electromagnetic pulses is attributed to the acceleration of emerging NS particles from the star's magnetic poles \cite{sahakian}. The magnetic field outside the star is of poloidal shape, and as the radiation comes out, they are channeled to narrow beams along the magnetic axis. However, pulsar's signals are in the form of pulses, which means the body magnetic axis and the rotation axis are tilted to each other. The misalignment of the magnetic axis and rotation axis results in the sweeping of light beams and has a lighthouse effect \cite{endean, wang,oliveira,lorimer}. If the magnetic axis is pointed towards the earth, it sweeps earth at regular intervals, 
and we observe pulses from pulsars. As energy is continuously drained out from the star through the magnetic poles, eventually, the star slows down, and the period of the pulsar changes over time \cite{lander1}.

Both rotation and magnetic fields play an essential role in the formation and emission of electromagnetic pulses from NS. The magnetic field in NS are thought to be flux frozen and does not change with rotation \cite{goldreich2,anderson,spruit}. In most of the problems discussed in the literature in connection to NS, the magnetic and rotation axis are taken to be aligned \cite{smith,aschenbach,michel,wada,cerutti}. For such a case, the problem becomes easier to solve. In ordinary NS this approximation can be permitted, as the magnetic field in such stars 
are such that it does not incorporate much change in the configuration of the star.

However, recent detection of pulsars with variable spin-down rates \cite{li}, attributed to their radiative precession \cite{melatos2}, and they wobble in their spin-down rates. Both the theoretical model and observation hints at the free precession of isolated NS \cite{lander1}. This free precession is only possible if there is an electromagnetic torque. This torque is provided by the super-strong magnetic field (rotating magnetic dipole \cite{davis,goldreich}) present in magnetars.
Few anomalous X-ray pulsars (AXPs) \cite{mereghetti,baykal1,baykal2} and soft gamma repeaters (SGR) \cite{Kouveliotou,kulkarni,murakami} shows such behaviors.
Measuring their pulse period over time yielded super-strong magnetic fields at their surfaces. The surface magnetic field are of 
the order of $10^{15}-10^{16}$ G \cite{paczynski,melatos,makishima}. They are now termed separately as magnetars \cite{duncan,thompson}.

%Magnetars are one of the most popular models to describe the central engine of gamma-ray bursts \cite{bernardini,cen,cui}. The rotational energy of magnetars is sufficient to power a significant fraction of magnetars. The prompt emission, the plateau, and the afterglow can be modeled successfully by magnetars. Such models' energy is provided either by the spin-down of magnetar (short GRB) or by accretion of matter on star surface (long GRB). 
To successfully model pulsars and magnetars, the assumption of the misalignment of rotation and magnetic axis is essential. For the case of a magnetar, what we have is an oblique rotator. 
%It is known that GW waves cannot be emitted by stationary axisymmetric stars but can be possible by magnetically non-axisymmetric deformed stars \cite{bonazzola,jones,ciolfi}. Therefore, to 
%successfully model GW and GRBs, one should address the problem of the oblique rotator.
In slowly rotating magnetars, the body symmetry axis is along the magnetic axis as the star is deformed due to the magnetic field. The rotating axis is inclined at an angle to the symmetry axis, resulting in an oblique rotator. The dynamics of the oblique rotator are still not well understood, and to address any problem of astrophysics of magnetars, we need to study this problem in detail. 

Previously, there had been some work on obliquely rotating magnetars. Mestel \& Takhar addressed the classical oblique rotator problem \cite{mestel}. They studied a general oblique rotator, which is connected to the main sequence star. Konno \cite{konno} addressed a problem
where both rotation and magnetic fields were present in the star; however, the two axes were aligned (not an oblique rotator). An oblique rotator problem for a magnetar was done later \cite{jones2,lander}, 
but the problem was treated classically.
In this paper, we address the problem of the oblique rotator in the general relativistic (GR) framework employing the perturbation technique. The frame-dragging (FD) effect can only be captured by the GR treatment, and they influence the path of the particle near a compact object.

As most of the magnetars have a period in the range of a few seconds (their rotational speed is not very fast), we assume that the distortion due to rotation is negligible, that is, the rotation does not cause the star to deviate from spherical symmetry. 
However, the deformation in the star's shape due to the magnetic field is significant, and we have an oblique rotator. The star is still axisymmetric but is tilted as shown in fig \ref{Mis}
In section II of this article, we first describe our system (the obliquely rotating star, the FD terms, the Einstein equation) and clearly state our assumptions. In section III, we present our results where we give the equation of motion (EOM), the geodesic equation and calculate the gravitational wave amplitude and the energy liberated by vacuum dipole magnetar. Finally, in section IV, we summarize our results and conclude them.

\section{The obliquely rotating star}

\begin{figure}
    %\vskip 0.2in
    \centering
    \includegraphics[scale=0.45]{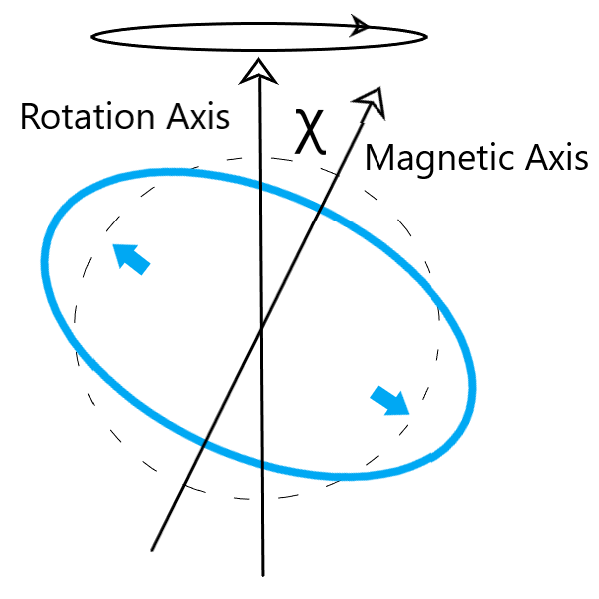}
    \caption{\small (Color online) The figure shows the misalignment of rotation and magnetic axis of an NS. Such kind of misalignment is mostly observed in pulsars, where the magnetic axis precesses around the rotation axis with an angle $\chi$. 
        Blue (solid line) shows magnetic deformation on the black (dashed line) background spherical star. For slowly rotating NS having a strong magnetic field (magnetar), there are no distortions due to rotation.}
    \label{Mis}
\end{figure}

We aim to study an axisymmetric star with the symmetry axis being the magnetic axis caused due to a strong magnetic field. The rotating axis does not coincide with the magnetic axis. This is the case for a rotating magnetar where the rotation axis and the magnetic axis are inclined to each other (fig \ref{Mis}).

The Lorentz force produced from the poloidal magnetic fields induces a bulge perpendicular to the magnetic axis and hence makes it axisymmetric with respect to the magnetic axis. The parameter that quantifies this distortion is
\begin{equation}
\epsilon_{B}\rightarrow\sqrt{\frac{B^2R^4}{GM^2}}.
\end{equation}
where $B$ is the magnetic field, $R$, and $M$ are the Tollmann-Oppenheimer-Volkof (TOV)  radius and mass of the spherically symmetric star \cite{tov} (having the same central density as the magnetically deformed star). 
Rotation, on the other hand, can be quantified by the parameter $\epsilon_{\Omega}\rightarrow\sqrt{\frac{\Omega^2R^3}{GM}}$ with $\Omega$ being the star's rotational frequency. These two parameters are the ratio of rotational and magnetic energies upon gravitational energies (dimensionless parameters), respectively.
If the rotation and magnetic axis had been aligned, the spherical star would deform to an axisymmetric star with both rotation and the magnetic axis being the symmetry axis. However for the misaligned case, if the angular velocity of the star is small and the magnetic field is strong (in case of magnetars), the symmetry axis would be the magnetic axis, and we would get a tilted axisymmetric star rotating (oblique rotator) about the rotation axis as shown in fig \ref{Mis}.

To solve the problem in GR perturbatively, the two important assumptions we make are as follows:

\begin{figure}[h!]
    %\vskip 0.2in
    \centering
    \begin{subfigure}[b]{0.55\textwidth}
        \includegraphics[scale=0.35]{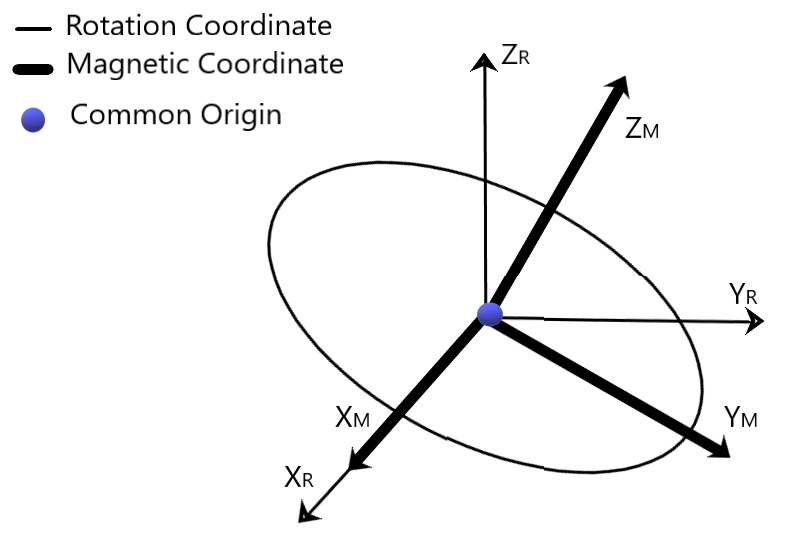}
        \caption{}
    \end{subfigure}
    
    \begin{subfigure}[b]{0.55\textwidth}
        \includegraphics[scale=0.25]{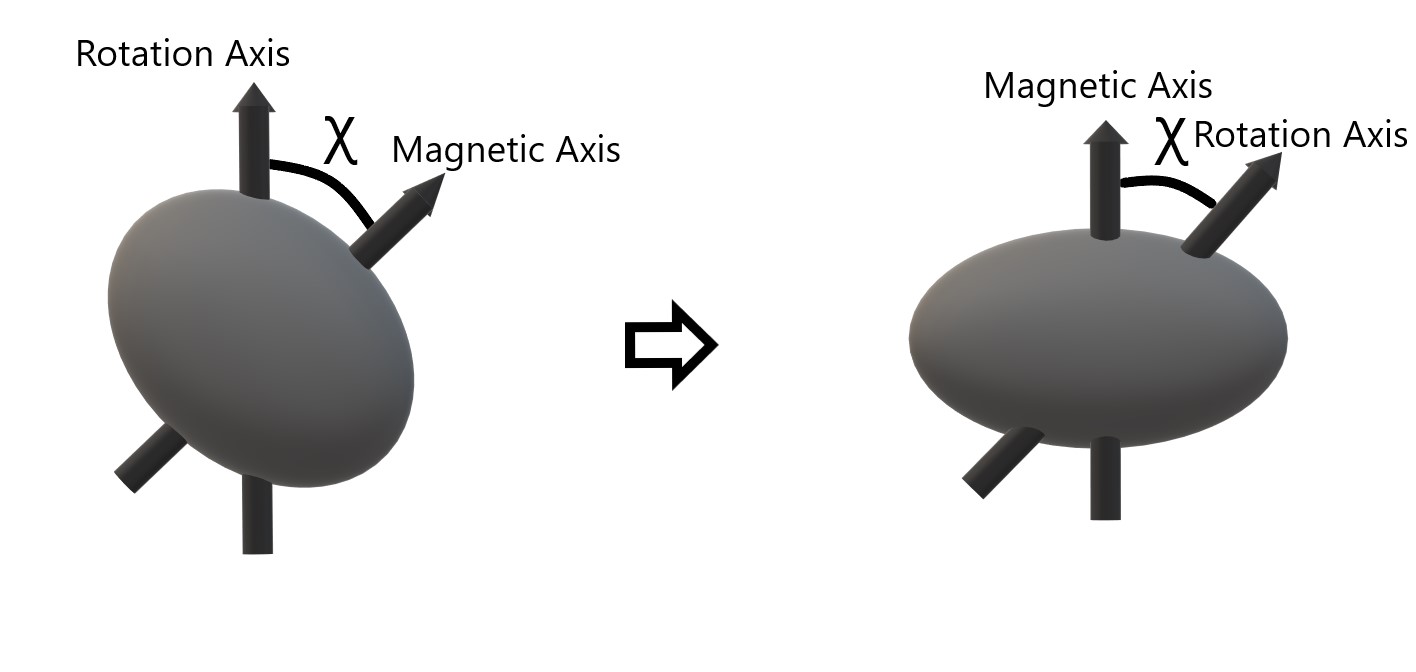}
        \caption{}
    \end{subfigure}
    
    \caption{\small(Color online) (a) Two sets of the coordinate system are shown (with the magnetic distortions taken into account) in the figure. The setup is such that rotational coordinates and magnetic coordinates share the same X-axis, and the $Z_M$ is precessing around the $Z_R$. As the distortion taken is purely due to the Lorentz force, the bulge is perpendicular to $Z_M$, and hence our system is axisymmetric about $Z_M$. (b) The transition from the rotational frame (the actual frame) to the magnetic frame. If we set up our 
        spherical coordinates around the magnetic axis instead of the rotation axis, the problem becomes simple.}
    \label{coord}
\end{figure}

\begin{itemize}
    \item The rotational frequency is small for the star such that, the star satisfies $R_{re}/R\approx 1$ and $R_{rp}/R\approx1$, where $R_{re}$ and $R_{rp}$ and $R$ are the equatorial, polar and TOV radius of the purely rotating star. On the other hand, the magnetic strength is large such that the deformation occurs only due to the magnetic field such that the star satisfies $R_e/R> 1$ and $R_p/R<1$, where $R_e$ and $R_p$ and $R$ are the equatorial, polar and TOV radius of the purely magnetized star respectively. The rotation and magnetic field can be compared in terms of the parameters $\epsilon_{\Omega}$ and $\epsilon_B$, which can be quantitatively related to each other as:
    \begin{equation}
    \epsilon_{\Omega}<<\epsilon_B<1
    \label{cond}
    \end{equation}
\end{itemize}

\begin{itemize}
    \item We will transform into a coordinate system defined by the magnetic axis. The rotation axis precess around it with frequency $\omega$ as shown in \cite{lander}, and is expressed as
    \begin{equation}
    \omega=\Omega\left(\frac{I_{zz}-I_{xx}}{I_{0}}\right)\cos{\chi}
    \label{ome}
    \end{equation}
    where $\chi$ is the inclination angle between the two axes, $I_{zz}$ and $I_{xx}$ are the moment of inertia for the z and x-axis of the perturbed star and $I_0$ is the moment of inertia for the unperturbed star.  Such a transformation is done by performing two sequence of rotational transformation. At first we define the two coordinate system, the rotational coordinate $\{X_R, Y_R, Z_R\}$, and the magnetic coordinate system $\{X_M, Y_M, Z_M\}$ where they share the same x-axis as shown in fig \ref{coord} (a). Then we can perform the first rotational transformation on the $X_M-Y_M$ plane about the $Z_M$ axis
        \begin{equation} 
        \begin{pmatrix}
        X_M  \\
        Y_M
        \end{pmatrix}=\begin{pmatrix}
        \cos\omega t & \sin\omega t &  \\
        -\sin\omega t & \cos\omega t &  \\
        \end{pmatrix}\begin{pmatrix}
        X_M  \\
        Y_M 
        \end{pmatrix},    
        \label{de1}
        \end{equation}  
such that $Y_M=\cos\omega t Y_M - \sin \omega t X_M$. 
But we also need to rotate our $Z_R-Y_R$ plane about the $X_R$ or the $X_M$ at an angle $\chi$, which is the second rotational transformation.
\begin{equation} 
\begin{pmatrix}
Y_R  \\
Z_R
\end{pmatrix}=\begin{pmatrix}
\cos\chi & -\sin\chi &  \\
\sin\chi & \cos\chi &  \\
\end{pmatrix}\begin{pmatrix}
Y_M  \\
Z_M 
\end{pmatrix}    
\label{de2}
\end{equation} 
 Combining the first and the second rotational transformation, we get
\begin{equation}
    Z_R=\cos\chi Z_M + \sin \chi (Y_M\cos\omega t-X_M\sin\omega t)
    \label{de}
\end{equation}
    
\end{itemize}
Equation (\ref{de}) shows that one can completely write the rotation axis $Z_R$ in terms of only the magnetic coordinate system. This setup helps in making our problem simpler \cite{lander} because, in this choice of coordinates, we retain our axisymmetry and also the magnetic field remains stationary. These two assumptions are pictorially explained in fig \ref{coord}(a) and \ref{coord}(b). 

\subsection{Frame-dragging} \label{Frame}
\begin{figure}[h!]
    \centering    
    \includegraphics[scale=0.4]{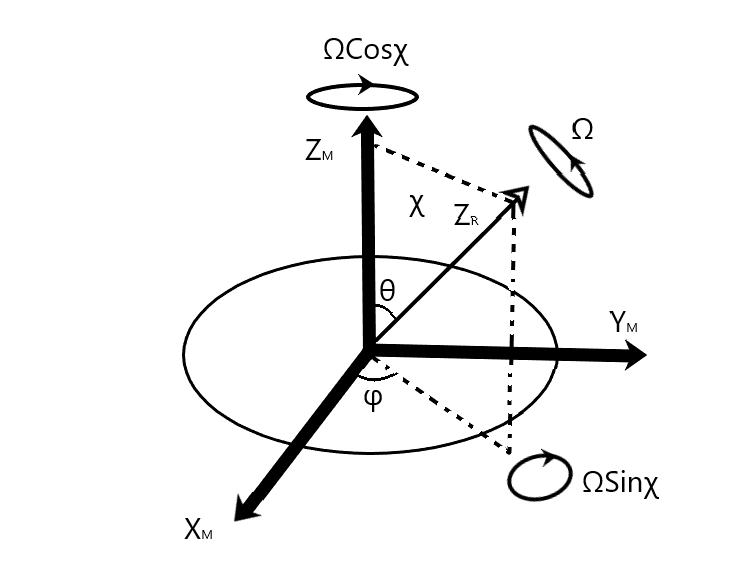}
    \caption{\small(Color Online) Decomposition of the angular velocities for a magnetar is shown.
        The rotation axis is tilted at an angle with the body axis, and we are in the transformed frame. The angular velocity now has components both along the symmetry axis and a plane perpendicular to it.}
    \label{drag1}
\end{figure}

Such a set up enables us to decompose the rotation axis vector $\Omega$ into two components, one along the body axis and another perpendicular to it (as shown in fig \ref{drag1}, and using eqn. (\ref{de}) the rotation can be written as
\begin{equation}
\vec{\Omega}=\omega \hat{Z}_M+\Omega \cos{\chi} \hat{Z}_M+\Omega \sin{\chi}(\hat{Y}_M \cos\omega t-\hat{X}_M \sin\omega t).
\label{rot}
\end{equation}
%The decomposition of $\vec{\Omega}$ is shown in fig \ref{drag1}. 

\begin{figure}[h!]
    \includegraphics[scale=0.23]{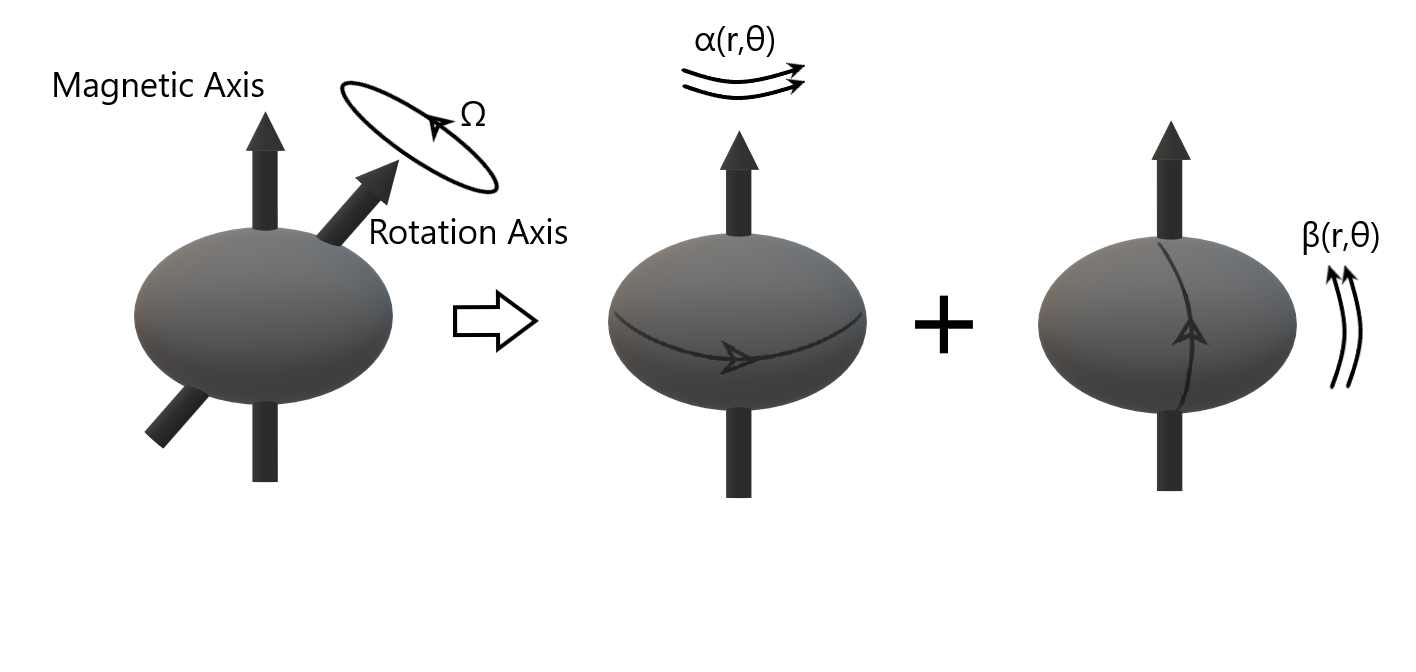}
    \caption{\small In the general relativistic framework, decomposition of rotational effect can be incorporated into the metric by adding two FD terms, 
        $\alpha(r,\theta)$ along the $\phi$ direction and $\beta(r,\theta)$ along the $\theta$ direction, and thus the overall FD for such a star is a 
        resultant of these two.}
    \label{drag2}
\end{figure}

From fig \ref{drag2}, one can see that any rotation for the magnetar can be decomposed into two independent rotations leading to two FD velocities,
\begin{itemize}
    \item $\alpha(r,\theta)$ which is along the $\phi$ direction and
    \item $\beta(r,\theta)$ which is along the $\theta$ direction.
\end{itemize}
In this framework, $r=Z_R$ and $\theta$ and $\phi$ are the angles shown in fig \ref{drag1}.
The FD terms are functions of $r$ and $\theta$ because we still have our axisymmetry preserved. For an axisymmetric star having rotation axis as the symmetry axis, only one FD term velocity arises around the symmetry axis. However, for an obliquely rotating axisymmetric star, we have two FD velocities one along the $\phi$ (azimuthal) direction and one along $\theta$ (polar) direction, as shown in fig \ref{drag2}.

\subsection{Equation of state}
To solve the TOV equation for stars, we need an equation of state (EoS) describing the properties of the matter of the star.
NS matter at the inner core is very dense, and they interact via the strong interaction. The degree of freedom at such densities is highly unpredictable and even disputed.
Therefore, to begin with, we assume that the degree of freedom is mainly neutron, proton, 
electron. The nuclear force carriers are assumed to be $\sigma$, $\omega$, and $\rho$. In this calculation we 
choose PLZ \cite{reinhard} and NL3 \cite{serot,glendenning}
parameter setting to describe the NM. 
The EoS is consistent with the recent astrophysical and nuclear constraint and can generate stars more massive than two solar mass. For the PLZ EoS, the radius of the star of mass $1.4$ solar mass is about $12$ km, which is also consistent with recent astrophysical constraints. Our analysis does not depend on the microphysics, 
and qualitative results remain
the same for any EoS (even for polytropic one). Therefore, we restrict our analysis with only these two EoS.

\subsection{The metric}
%\subsection{Metric Elements}
We start with a spherically symmetric non-magnetic static star whose metric is given by
\begin{equation}
ds^{2}=-e^{\nu(r)}dt^{2}+e^{\lambda(r)}dr^{2}+r^2d\theta^2+r^2\sin^2{\theta}d\phi^2.
\end{equation}
To include rotation in our metric, we incorporate the two FD terms $\alpha$ and $\beta$ as was discussed in section \ref{Frame}. This rotation, however, does retain the spherical symmetry of the star (for slow rotation the deviation from spherical symmetry is neglected). The $\alpha(r)$ FD takes place in the direction of $\phi$ while the $\beta(r)$ FD takes place in the direction of $\theta$. Thus the metric becomes 
\begin{align}
    ds^{2}&=-e^{\nu(r)}dt^{2}+e^{\lambda(r)}dr^{2}+r^2\Big[d\theta-(\sin\chi)\beta(r)dt\Big]^2 \nonumber \\
    &+r^2\sin^2{\theta}\Big[d\phi-(\cos\chi)\alpha(r)dt\Big]^2,
\end{align}
where $\chi$ is the misalignment angle. This metric is valid for slow rotation, as we do not include deformation due to rotation in our metric coefficients.
In all our calculation from now, we work in the geometrized unit, that is $G=c=1$. The poloidal axisymmetric magnetic field $B$ of the order $O(\epsilon_{B})$ get added upon this background metric. The magnetic field acts as a perturbation over this spherical, slowly rotating star and causes a bulge along the equator and contraction along the poles, as evident in figs. (\ref{coord}) and (\ref{drag2}). The explicit form of the magnetic field would be defined in the next section.
We still have our axisymmetry preserved if we work in our chosen magnetic coordinate frame. 
Thus the metric up to order $O(\epsilon_{B}^2)$ is
\begin{align}
    ds^{2}&=-e^{\nu(r)}(1+2h(r,\theta))dt^{2}+e^{\lambda(r)}\Big(1+\frac{2e^{\lambda(r)}m(r,\theta)}{r}\Big)dr^{2}\nonumber \\
    &+r^2(1+2k(r,\theta))\Big[\Big(d\theta-(\sin\chi)\beta(r,\theta)dt\Big)^2\nonumber \\
    &+\sin{\theta}^2\Big(d\phi-(\cos\chi)\alpha(r,\theta)dt\Big)^2\Big].
    \label{metric1}
\end{align}

In order to see the effect of rotations, we expand both $\alpha(r,\theta)$ and $\beta(r,\theta)$ up to order $O(\epsilon_{\Omega}\epsilon_{B}^2)$,
\begin{eqnarray}
\alpha(r,\theta)=\alpha_0(r,\theta)+\alpha_1(r,\theta)\\
\beta(r,\theta)=\beta_0(r,\theta)+\beta_1(r,\theta)
\end{eqnarray}

We understand that the EOM for $\alpha_0$ (the FD frequency) does not include any source term \cite{hartle}. As the order of $\alpha_0$ and $\beta_0$ is the same $O(\epsilon_{\Omega})$; thus, $\beta_0$ will also not include any source term. With slow rotation approximation, we only keep terms till $O(\epsilon_{\Omega})$. However, as we want to study magnetars, we keep terms of the order $O(\epsilon_{\Omega}\epsilon_{B})$ and $O(\epsilon_{\Omega}\epsilon_{B}^2)$. Higher orders, $O(\epsilon_{\Omega}^2\epsilon_{B}^2)$ and above, are neglected because both $\epsilon_{\Omega}$ and $\epsilon_B$ are less than unity. Hence EOM for $\alpha_1$ and $\beta_1$ is necessary to see the coupling between rotation and magnetic field. 
The orders of all the unknown functions are as follows
\begin{eqnarray}
(h(r,\theta),m(r,\theta),k(r,\theta))\rightarrow O(\epsilon_{B}^2) \\
(\alpha_0(r,\theta),\beta_0(r,\theta))\rightarrow O(\epsilon_{\Omega}) \\
(\alpha_1(r,\theta),\beta_1(r,\theta))\rightarrow O(\epsilon_{\Omega}\epsilon_{B}^2). 
\end{eqnarray}

Now in order to decouple $r$ and $\theta$, we expand these unknown functions in terms of Legendre polynomial,
\begin{eqnarray}
\label{exp1}
h(r,\theta)=h_0(r)P_0(\cos{\theta})+h_2(r)P_2(\cos{\theta}) \\
m(r,\theta)=m_0(r)P_0(\cos{\theta})+m_2(r)P_2(\cos{\theta}) \\
k(r,\theta)=k_0(r)P_0(\cos{\theta})+k_2(r)P_2(\cos{\theta}) 
\end{eqnarray}
and
\begin{eqnarray}
\alpha_0(r,\theta)=\alpha_0(r)A(\theta)\\
\alpha_1(r,\theta)=\alpha_1(r)A(\theta) \\\
\beta_0(r,\theta)=\beta_0(r)A(\theta) \\
\beta_1(r,\theta)=\beta_1(r)A(\theta)
\label{exp2}
\end{eqnarray}
where $A(\theta)\equiv\Big(-\frac{1}{\sin\theta}\frac{dP_1(\cos{\theta})}{d\theta}\Big)$. Substituting equation (\ref{exp1})-(\ref{exp2}) in our metric (\ref{metric1}) and putting $P_0(\cos\theta)=1$, we get our final metric line element.
Form equation (\ref{metric1}), we see that we have two cross terms $g_{t\theta}$ and $g_{t\phi}$ in the metric. The $g_{t\phi}$ term is present in any axisymmetry metric, however, the  presence of a $g_{t\theta}$ term is inherent when there is a misalignment and the rotation axis is not the symmetry axis.
However such a term will be weighted by a function $f(\chi)g_{t\theta}$ such that $\lim\limits_{\chi\rightarrow0}f(\chi)=0$. Here our simple choice 
of $f(\chi)\equiv\sin\chi$. We also want to see the case of $\chi=\frac{\pi}{2}$ misalignment, where the rotation and magnetic axis are perpendicular to each other. The $g_{t\phi}$ term is weighted as $g(\chi)g_{t\phi}$ such that $\lim\limits_{\chi\rightarrow0}g(\chi)=1$, so we choose $g(\chi)\equiv\cos\chi$. Our choice of these functions is motivated from equation (\ref{rot}), as we see that in the limit $\chi\rightarrow0$, we get back the metric of a star whose rotation and magnetic axis is aligned. Using the metric given by equation 
(\ref{metric1}) we can 
calculate the Christoffel's symbols, the Riemann tensor and finally the Ricci tensor in order to calculate the Einstein tensor $G_{\mu\nu}$.

\subsection{Stress-Energy Tensor} \label{Mag}
We calculate the stress-energy tensor assuming perfect fluid inside the star. Thus the matter part is given by
\begin{equation}
\label{stress}
T_{\mu\nu}^{Matter}=(\epsilon+P)U_{\mu}U_{\nu}+Pg_{\mu\nu}
\end{equation}
where $U^{\mu}=\frac{dx^{\mu}}{d\tau}$
are the four velocities of the fluid particles inside the star. 
The presence of nonzero $U^\theta$ shows misalignment. The pressure and density for the matter can be 
%$$G_{\mu\nu}=T_{\mu\nu}$$
expanded in terms of Legendre polynomials
\begin{align}
&\epsilon(r,\theta)=\epsilon_{0}(r)+(\epsilon_{02}(r)+\epsilon_{22}(r)P_{2}(\cos\theta)) \\
&p(r,\theta)=p_{0}(r)+(p_{02}(r)+p_{22}(r)P_{2}(\cos\theta)).
\end{align}
We use a one-parameter EoS in order to find a relation between $p$ and $\epsilon$ where $p=p(\epsilon)$. Observing that $\epsilon_{02}$ and $\epsilon_{22}$ are small changes in $\epsilon$ around $\epsilon_{0}$, using partial derivative ${dp}=\frac{\partial p}{\partial \epsilon}{d\epsilon}$ we have,
\begin{equation}
\epsilon(r,\theta)=\epsilon_{0}(r)+\frac{\partial \epsilon}{\partial p}(p_{02}(r)+p_{22}(r)P_{2}(\cos\theta)).
\end{equation}
Provided the energy density, pressure and four-velocity, $T_{\mu\nu}^{Matter}$ can be calculated from equation (\ref{stress}). The stress-energy tensor for the magnetic field is written as 
\begin{equation}
T_{\mu\nu}^{EM}=\Big[g^{\alpha\beta} F_{\mu\beta}F_{\nu\alpha}-\frac{1}{4}g_{\mu\nu}g^{\sigma\alpha}g^{\delta\beta}F_{\alpha\beta}F_{\sigma\delta}\Big],
\end{equation}
where the electromagnetic field tensor is defined as $
F_{\mu\nu}\equiv\partial_{\mu}A_{\nu}-\partial_{\nu}A_{\mu}$
and $A_{\mu}$ is the electromagnetic four potential. We have an axisymmetric poloidal magnetic field created by a four current $J_\mu=(0,0,0,J_\phi)
$. The electromagnetic four potential is
\begin{equation}
\label{mag}
A_{\mu}=(A_t,0,0,A_\phi).
\end{equation} 
where $A_t$ is the rotationally induced parameter due to the presence of poloidal magnetic field potential $A_\phi$ \cite{konno,christian}. The current $J_\phi$ along with the star's rotational component $\Omega\cos{\chi}$ in the direction of the axisymmetry are the source for the potential $A_{\mu}$.
The three functions, $A_t(r,\theta)$, $A_{\phi}(r,\theta)$ and $J_\phi(r,\theta)$ can be  expanded in terms of 
Legendre polynomials
\begin{eqnarray}
A_\phi(r,\theta)=a_\phi(r)(\sin\theta\frac{dP_1(\cos\theta)}{d\theta}) \\
\label{mag3}
A_t(r,\theta)=a_{t0}(r)+a_{t2}(r)P_{2}(\cos\theta)\\
J_\phi(r,\theta)=j_\phi(r)(\sin\theta\frac{dP_1(\cos\theta)}{d\theta})
\end{eqnarray}
The two potentials, $A_\phi$ and $A_t$ will satisfy the Maxwell equation given by
\begin{align}
e^{-\lambda}\frac{d^2a_\phi}{dr^2}+\frac{1}{2}(\nu'-\lambda')e^{-\lambda}\frac{da_\phi}{dr}-\frac{2}{r^2}a_\phi=-4\pi j_\phi
\end{align}

%\begin{widetext}
\begin{equation}
\frac{d^2a_{t0}(r)}{dr^2}=\frac{1}{6r^2}\Big[8e^\lambda a_\phi\alpha_0(r)\cos{\chi}-12r\frac{da_{t0}(r)}{dr}+4r\frac{da_\phi}{dr}\Big(\alpha_0(r)\cos\chi(2+r\lambda')+r\frac{d\alpha_0(r)}{dr}\cos\chi\Big)\Big]
\label{at0}
\end{equation}
\begin{equation}
    \frac{d^2a_{t2}(r)}{dr^2}=\frac{1}{r^2}\Big[6e^\lambda a_{t2}(r)-2r\frac{da_{t2}(r)}{dr}+2r\frac{da_\phi}{dr}\Big(\alpha_0(r)\cos\chi(2+r\lambda')+r\frac{d\alpha_0(r)}{dr}
    \cos\chi\Big)\Big].
    \label{at2}
    \end{equation}
%\end{widetext}
In the vicinity of the star center, the potential $a_\phi$ and the four-current is given by \cite{konno}
\begin{eqnarray}
a_\phi\simeq a_0r^2+O(r^4) \\
J_{\phi}=c_0 r^2(\epsilon_0(r)+p_0(r))
\end{eqnarray}
where $a_0$ and $c_0$ are constants. The value of $a_0$ and $c_0$ determines the strength of the magnetic field B. In our calculation, we have taken values of $a_0$ and $c_0$ to be same.
The order of the unknown functions that is the pressure, density, the magnetic potential, and the rotationally induced potentials are

\begin{eqnarray}
(p_{02},p_{22},\epsilon_{02},\epsilon_{22})\rightarrow O(\epsilon_B^2) \\
a_{\phi}\rightarrow O(\epsilon_{B}) \\
(a_{t0}(r),a_{t2}(r))\rightarrow O(\epsilon_{\Omega}\epsilon_{B}).
\end{eqnarray}

Finally, we solve the Einstein's equation order wise.
Our main aim is to find the FD EOM that is given by $G_{t\phi}$ and $G_{t\theta}$. 

%\begin{widetext}
\section{Results}
In this section we first calculate the EOM solving the Einstein equation. 
\subsection{Equation of motion}
The EOM will be calculated order wise. The $\alpha$ EOM are
\subsubsection{EOM $G_{t\phi}=T_{t\phi}$}
\begin{itemize}
    \item \underline{$R_{t\phi}$ up to order $O(\epsilon_{\Omega})$}
    \begin{align}
        &\frac{d^2\alpha_0(r)}{dr^2}-\frac{1}{2r}e^{-\lambda}\Big[r(\lambda'+\nu')-8\Big]\frac{d\alpha_0(r)}{dr}+2(p_{0}+\epsilon_{0})\alpha_0(r)=0
        \label{EOM1}
    \end{align}

    \item \underline{$R_{t\phi}$ up to order $O(\epsilon_{\Omega}\epsilon_{B}^2)$}
    \begin{align}
    &\frac{d^2\alpha_{1}(r)}{dr^2}-\frac{1}{2r}\Big[r(\lambda'+\nu')-8\Big]\frac{d\alpha_{1}(r)}{dr}+\frac{2}{r}(\lambda'+\nu')\alpha_{1}(r)=-(S_0-\frac{S_2}{5})
    \label{EOM2}
    \end{align}    
\end{itemize}
where, $S_0$ and $S_2$ are sourse terms given by 
    \begin{align}
    S_0&={\alpha}'_0(r)\Big[-h_0-\frac{e^\lambda m_0}{r}\Big]'-\frac{4e^\lambda(\nu'+\lambda')\alpha_0(r)m_0}{r^2}-4e^\lambda\bar{\alpha}_0(r)(\rho_{20}+p_{20})
    +\frac{16e^\lambda a_\phi^2\alpha_0(r)}{3r^4}+\frac{8(a_\phi')^2\alpha_0(r)}{3r^2}\nonumber\\
    &+\frac{8e^\lambda a_\phi a_{t2}}{r^4}-\frac{4a_\phi'a_{t0}'}{r^2}
    \label{s0}
    \end{align}
%\end{widetext}
%\begin{widetext}
    \begin{align}
    S_2&=\alpha_0'(r)\Big[4k_2-h_2-\frac{e^\lambda m_2}{r}\Big]'-\frac{4e^\lambda(\nu'+\lambda')\alpha_0(r)m_2}{r^2}
    -4e^\lambda\alpha_0(r)(\rho_{20}+p_{20})+\frac{32e^\lambda a_\phi^2\alpha_0(r)}{3r^4}-\frac{8(a_\phi')^2\alpha_0(r)}{3r^2}\nonumber\\
    &+\frac{16e^\lambda a_\phi a_{t2}}{r^4}-\frac{4a_\phi'a_{t2}'}{r^2}
    \label{s2}
    \end{align}

\subsubsection{EOM $G_{t\theta}=T_{t\theta}$}

\begin{itemize}
    \item \underline{$R_{t\theta}$ up to order $O(\epsilon_{\Omega})$}
    \begin{align}
    &\frac{d^2\beta_0(r)}{dr^2}-\frac{1}{2r}e^{-\lambda}\Big[r(\lambda'+\nu')-8\Big]\frac{d\beta_0(r)}{dr}+2(p_{0}+\epsilon_{0})\beta_0(r)=0
    \label{EOM3}
    \end{align}
    \item \underline{$R_{t\theta}$ up to order $O(\epsilon_{\Omega}\epsilon_{B}^2)$}
    \begin{align}
    &\frac{d^2\beta_{1}(r)}{dr^2}-\frac{1}{2r}\Big[r(\lambda'+\nu')-8\Big]\frac{d\beta_{1}(r)}{dr}+\frac{2}{r}(\lambda'+\nu')\beta_{1}(r)=\frac{R_0+R_2}{2}
    \label{EOM4}
    \end{align}
    
\end{itemize}
where, $R_0$ and $R_2$ are sourse terms given by
    \begin{align}
    R_0&=4e^{\lambda}p_{02}(r)\frac{(p'_0(r)+\epsilon'_{0}(r))}{p'_{0}(r)}\bar{\beta}_0(r)+\frac{2}{r^2}e^{\lambda}\Big[m_0(r)(r\nu'(r)-7)-r\frac{dm_0(r)}{dr}\Big]
    \frac{d\beta_0(r)}{dr}-\frac{32}{3r^4}e^{\lambda}a_{\phi}^2(r)\beta_0(r)\nonumber\\
    &-\frac{4}{r}e^{\lambda}m_0(r)\frac{d^2\beta_0(r)}{dr^2}+2\frac{dh_0(r)}{dr}\frac{d\beta_0(r)}{dr}
    \label{r0}
    \end{align}
%\end{widetext}
%\begin{widetext}
    \begin{align}
    R_2&=4e^{\lambda}p_{22}(r)\frac{(p'_0(r)+\epsilon'_{0}(r))}{p'_{0}(r)}\bar{\beta}_0(r)+24\frac{e^{2\lambda}}{r^3}m_{2}(r)\beta_0(r)-2\Big[h'_2(r)-4k'_2(r)\Big]\frac{d\beta_0(r)}{dr}+\frac{16}{r^2}e^{\lambda}k_2(r)\beta_0(r)\nonumber\\
    &-\frac{2}{r}e^\lambda\frac{d\beta_0(r)}{dr}\frac{dm_2(r)}{dr}+\frac{2}{r^2}e^\lambda m_2(r)\Big[r\nu'-7\Big]\frac{d\beta_0(r)}{dr}-\frac{4}{r}e^\lambda m_2(r)\frac{d^2\beta_0(r)}{dr^2}-\frac{18}{r^4}e^\lambda\Big[rh_2(r)+e^\lambda m_2(r)\Big]\beta_0(r) 
    \label{r2}
    \end{align}
    
%    \end{widetext}
Equation (\ref{EOM1}) is Hartle's FD EOM, which has no source term and hence no effect of the magnetic field up to order $O(\epsilon_{\Omega})$. 
Thus up to $O(\epsilon_{\Omega}\epsilon_B^2) $, equation (\ref{EOM2}) gives a contribution of the magnetic field to the rotational effect. Similarly,
equation (\ref{EOM3}) also has no contribution from the magnetic field. However, EOM for $\beta_1$ has source terms, but we observe that rotationally induced electric field terms are absent here. The nature of such FD terms is seen after numerically solving the EOM for $\alpha$ and 
$\beta$ and doing a comparative study.

\begin{figure}[ht!]
    %\vskip 0.2in
    \centering
    \begin{subfigure}[b]{0.65\textwidth}
        \includegraphics[scale=0.65]{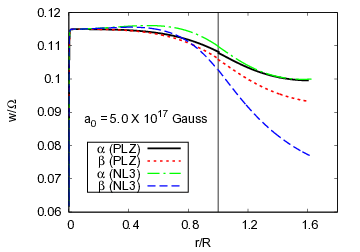}
        \caption{}
    \end{subfigure}
    
    \begin{subfigure}[b]{0.65\textwidth}
        \includegraphics[scale=0.65]{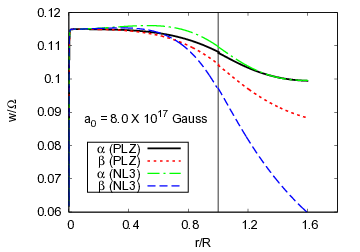}
        \caption{}
    \end{subfigure}
    \caption{\small(Color online) FD frequencies as a function of radial distance are shown in the figure for NL3 and PLZ EoS. The two plots are for two different central magnetic field (a) $a_0$ = $5.0\times10^{17}$ G  
        (b) $a_0$= $8.0\times10^{17}$ G at the star's centre.
        The FD is normalized by the total rotational velocity of the star, and the radius of the star normalizes the radial distance. As the magnetic field 
        changes, the FD velocities hit the surface with different values, where the $\alpha$ FD velocity falls faster in (a) than in (b).}
    \label{PLZ2013}
\end{figure}

In our calculation, we have used a poloidal magnetic field as defined in equation (\ref{mag}). We have shown our results for two central magnetic fields values, $5 \times 10^{17} G$ 
and $8 \times 10^{17} G$. Such values are well within the conservative choices of magnetic field \cite{bocquet,cardall,mallick-deform,debarati,schramm,veronica}, and are 
expected to be present in magnetars whose surface magnetic 
field are of the order of $10^{14} - 10^{15} G$. Using such magnetic fields in a slowly rotating star we find both the FD velocities $\alpha$, and $\beta$ has different nature of evolution. 
For  NL3 EoS starting from the center of the star, both $\alpha$ and $\beta$ gradually increases and reaches a maximum at around 0.5 times of the star's radius (fig \ref{PLZ2013}). 
Both of them gradually decreases as we get towards the surface of the star. The FD $\beta$ is initially of the same magnitude of the $\alpha$ but falls more rapidly than $\alpha$. This is because of the presence of the rotationally 
induced electric field in the source terms of $\alpha$ (differential equation (\ref{s0}) and (\ref{s2})), but is absent in the source term of $\beta$ (differential 
equation (\ref{r0}) and (\ref{r2})). For PLZ EoS, the nature of the curve is similar; however, the curves ($\alpha$ and $\beta$) starts to fall from the center. However, in both cases, $\beta$ is always smaller than $\alpha$. The nature of FD remains the same for two EoS, and the only difference in the plots is due to the difference in stiffness of EoS. The NL3 is steeper than PLZ, which means that for the same energy density values, NL3 produces higher pressure than PLZ. Stars having NL3 EoS, the pressure and density fall faster with radial distance, the driving force gets weaker, and thus FD velocities fall faster.

This rotationally induced electric field occurs because of the rotation around the direction of the magnetic field, that is 
$A_t\rightarrow O(\Omega)\times A_\phi$ and hence $A_t$ and $A_\phi$ couples in the Einstein equation $G_{t\phi}=T_{t\phi}$. However because of the absence of 
$A_\theta$ potential, the $A_t$ terms doesn't appear in the equation $G_{t\theta}=T_{t\theta}$. This effects the $G_{t\phi}$ equation and acts as a driving force for its evolution, and thus it hits the surface with a higher value than the $G_{t\theta}$ FD where this extra driving force is not present. If the magnetic field is lower than $10^{15}$ G, then this separation is not significant.
As we increase the magnetic field, the two FD separates more, and $\beta$ starts falling faster than $\alpha$ (fig \ref{PLZ2013}(a) and \ref{PLZ2013}(b)).  The FD velocity along the azimuthal direction ($\phi$) is greater than the FD velocity along the polar ($\theta$) direction due to the rotation induced electric field. Therefore, particles are dragged along the azimuthal direction, but their orbit is not planar. The particle also moves along the polar direction.

\subsection{Geodesic of a particle near the star}
The FD frequencies have a massive impact on defining the geodesic of particles moving in and around the star. Therefore, 
it is interesting to see how a test particle behaves around such an obliquely rotating magnetar. The motion is governed by the geodesics constructed around such stars. 
The geodesic equation is given by
\begin{equation}
\frac{d^2x^\mu}{dt^2}=-\Gamma_{\alpha\beta}^\mu\frac{dx^\alpha}{dt}\frac{dx^\beta}{dt}-\frac{1}{\frac{dt}{d\tau}}\frac{d}{dt}\Big[\frac{dt}{d\tau}\Big]\frac{dx^\mu}{dt}.
\label{geo} 
\end{equation}
Using this equation, we calculate the geodesic equation for $r(t)$, $\theta(t)$ and $\phi(t)$, which gives us the three dimensional motion of a test 
particle around our star. We calculate them in perturbative fashion order wise.

%\begin{widetext}
\subsubsection{Geodesic Equation up to order O($\epsilon_{\Omega}$)}
%We calculate the geodesic equation up-to order $O(\epsilon_{\Omega})$

\begin{itemize}
    \item \textbf{r-Geodesic up to order $O(\epsilon_{\Omega})$}
        \begin{align}
        r''(t)=-\frac{e^{-\lambda}}{2}\Big[-2r\theta'(t)^2-2r\sin^2\theta\phi'(t)^2+e^\lambda r'(t)^2\lambda'+e^\nu\nu'\Big]-e^{-\lambda}r\Big[\sin\chi\Big(2\beta_0+r\beta_0'\Big)\theta'(t) \nonumber \\
        +\cos\chi\Big(\sin^2\theta(2\alpha_0+r\alpha'_0)\Big)\phi'(t)\Big]
        \label{eqr}
        \end{align}
%    \end{widetext}
    \item \textbf{$\theta$-Geodesic upto order $O(\epsilon_{\Omega})$}
%    \begin{widetext}
        \begin{align}
        \theta''(t)=\Big[-\frac{2r'(t)\theta'(t)}{r}+\frac{\sin 2\theta \phi'(t)^2}{2}\Big]+\sin\chi\Big[\frac{2\beta_0r'(t)}{r}+r'(t)\frac{d\beta_0}{dr}-\beta_0r'(t)\frac{d\nu}{dr}\Big]-\cos\chi\Big[\sin 2\theta\alpha_0\phi'(t)\Big]
        \label{eqtheta}
        \end{align}
%    \end{widetext}
    \item \textbf{$\phi$-Geodesic upto order $O(\epsilon_{\Omega})$}
%    \begin{widetext}
        \begin{align}
        \phi''(t)=\Big[\frac{-2(r'(t)+r\cot\theta\theta'(t))\phi'(t)}{r}\Big]+\frac{\cos\chi}{r}\Big[2r\cot\theta\alpha_0\theta'(t)
        +r'(t)(\alpha_0(2-r\frac{d\nu}{dr})+r\frac{d\alpha_0}{dr})\Big]
        \label{eqphi}
        \end{align}
\end{itemize}
%\end{widetext}

The first terms in these three equations we know are for the geodesic of a particle around a spherically symmetric star. The rest of the terms comes due to the rotation of the star till order $O(\epsilon_{\Omega})$. However, in this order, we do not have a contribution from the magnetic field of the star. For that, we have to 
calculate the geodesics to the next order in the magnetic field.

\subsection{Geodesic Equation of order O($\epsilon_{\Omega}\epsilon_B^2$)}
\begin{itemize}
    \item \textbf{r-Geodesic of order $O(\epsilon_{\Omega}\epsilon_B^2)$}\\\\
    The radial equation tells us how the particle trajectory changes as we radially move outwards. However, in our analysis, we only analyze a particle in some particular orbit at a fixed radial distance. Therefore, the big radial equation is given in Appendix A.
    
    \item \textbf{$\theta$-Geodesic of order $O(\epsilon_{\Omega}\epsilon_B^2)$}
%    \begin{widetext}
        \begin{align}
        \theta''(t)&=-\frac{A}{r^3}\Big[-e^\nu r\cot\theta h_2+e^{2\lambda}\cot\theta m_2r'(t)^2+r^3\Big(\frac{dk_2}{dr}r'(t)\theta'(t)-\cot\theta k_2\theta'(t)^2+\frac{\sin 2\theta k_2\phi'(t)^2}{2}\Big)\Big]\nonumber\\ &+\Big[\sin\chi\Big(A\cot\theta h_2\beta_0\theta'(t)+\frac{r'(t)}{2r}\Big(r\beta_0\Bigg(-4\frac{dh_0}{dr}+A\left(-\frac{dh_2}{dr}
        +\frac{dk_2}{dr}\right)\Bigg)\nonumber\\
        &+2r\frac{d\beta_1}{dr}+\beta_1(4-2r\frac{d\nu}{dr})\Big)\Big)-\cos\chi\frac{\sin2\theta}{2}\Big(2\alpha_1-A k_2\alpha_0\Big)\phi'(t)\Big]
        \label{theta}
        \end{align}
%    \end{widetext}
    \item \textbf{$\phi$-Geodesic of order $O(\epsilon_{\Omega}\epsilon_B^2)$}
%    \begin{widetext}
        \begin{align}
        \phi''(t)&=-\frac{A}{2}\Big(\frac{dk_2}{dr}r'(t)-2\cot\theta k_2\theta'(t)\Big)\phi'(t)+\Big[\frac{\cos\chi}{4}\Big(r'(t)(2\alpha_0\Bigg(-4\frac{dh_0}{dr}-A\left(\frac{dh_2}{dr}-\frac{dk_2}{dr}\right)\Bigg)\nonumber\\
        &+4\frac{d\alpha_1}{dr})+2B(h_2-k_2)\alpha_0\theta'(t)+\frac{4}{r}(\alpha_1(2r\cot\theta\theta'(t)+r'(t)(2-r\frac{d\nu}{dr})))\Big)\Big]
        \label{phi}
        \end{align}
%    \end{widetext}
where, $A=1+3\cos2\theta$ and $B=\frac{5\cos\theta + 3\cos 3\theta}{\sin\theta}$.
\end{itemize} 
%\end{widetext}
In this order, the geodesics have a contribution from the magnetic field and thus acts as the driving force for the differential equation of $r(t),\theta(t),\phi(t)$. These geodesic equations are solved numerically and the particles orbits are plotted below.

\begin{figure}[h!]
    %\vskip 0.2in
    \centering
    \begin{subfigure}[b]{0.5\textwidth}
        \includegraphics[scale=0.7]{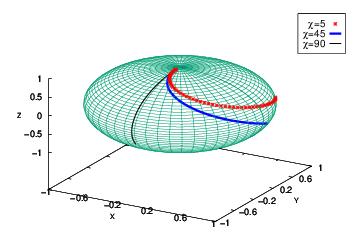}
        \caption{}
    \end{subfigure}
    
    \begin{subfigure}[b]{0.5\textwidth}
        \includegraphics[scale=0.7]{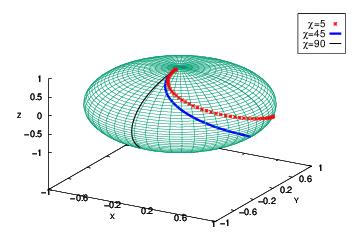}
        \caption{}
    \end{subfigure}
    \caption{\small(Color online) Geodesic of a particle near the surface of the star is shown in the figure for PLZ EoS. It has been plotted for two different central magnetic field (a) $a_0$= $5.0\times10^{17}$ G 
        (b) $a_0$= $8.0\times10^{17}$ G at the center of the star.
        We have plotted the geodesic for misalignment angle $\chi=5,45,90$ degrees. The three angles give different paths for particles around the star. As the angle increases, 
        the $\beta$ FD dominates and the geodesic shifts. As the magnetic field increases from (a) to (b), all of the three misaligned geodesic shifts.}
    \label{GNL6013}
\end{figure}

We have plotted the geodesic of a massless test particle on the star's surface, taking the radius to be constant and varying $\theta$ and $\phi$ as a function of time. Geodesics are shown for stars having two different magnetic fields modeled with PLZ EoS. The results for 
NL3 EoS are almost the same and are not shown here.
A test mass starting 
from $\theta=0$ and $\phi=0$ has been plotted from $t=0$ to $t=t_0$. All the other functions have been taken at $r=R$. 
Assuming the particles starts from rest at the pole defines the initial conditions $\phi(t)=\phi_0$ and $\theta(t)=0$ and the initial velocities, 
$\theta'(t)=0$ and $\phi'(t)=0$. Thus, the movement of the particle is only because of the FD term ($\alpha$ and $\beta$) generated by the rotation and magnetic field of the star. 

In general, the geodesics on a sphere are great circles. This could be seen from the first two terms of equation (\ref{eqr}),(\ref{eqtheta}),(\ref{eqphi}). 
However, in our calculation, both rotation and magnetic fields are present. 
$\chi=0^{\circ}$ is an aligned rotator and the geodesic are are circles therefore, we do not show then here.
Let us start with the situation when $\chi=5^{\circ}$, that there is very small misalignment. In this case, initially the $\alpha$ FD is dominant; thus, a particle around the star's surface is dragged mostly towards the azimuthal direction. However, even if we leave the particle with zero velocity ($\theta'=\phi'=0$), with time, first $\phi'$ will be non zero (even for order $O(\epsilon_{\Omega})$, last term of eqn \ref{eqphi}) which will then induce a non zero $\theta'$. Therefore, the particle will not only move azimuthally with time but also would move down along polar ($\theta$) direction. The red curve in fig \ref{GNL6013} shows the geodesic of a particle of an aligned rotator. 

For large misalignment angle ($\chi=45^{\circ}$), then we have an oblique rotator with body symmetry axis being the magnetic axis. Here, both $\alpha$ and $\beta$ terms are present, and both these terms
would contribute. All the terms in geodesic equation (\ref{eqtheta}, \ref{eqphi}, \ref{theta} and \ref{phi}) other than $r'$ term will contribute towards defining the trajectory of the test particle. As a result, both $\phi'$ and $\theta'$ would pick non zero value quickly, and the trajectory of the particle move down along $\theta$ faster.
The particle now feels a resultant effect from both the FDs.

For $\chi=90^{\circ}$, the rotation and magnetic axis are mutually perpendicular. One does not affect the other. For such a case, we only have a contribution from $\beta$. For such a situation, if we leave the particle with zero velocity ($\theta'=\phi'=0$), then for the order $O(\epsilon_{\Omega})$ all the terms are zero. Only in the order $O(\epsilon_{\Omega}\epsilon^2_B)$ the $\theta$ term survives, and therefore the trajectory of the particle is only along the $\theta$ direction. The trajectory of the particle is a circle over the sphere.
As we increase our magnetic field, the driving force for $\alpha$  and $\beta$ increases, and thus, they reach a higher value, thus pushing the geodesic downwards (fig \ref{GNL6013}(b)).

\begin{figure}[h!]
    %\vskip 0.2in
        \centering
    \begin{subfigure}[b]{0.3\textwidth}
        \includegraphics[scale=0.35]{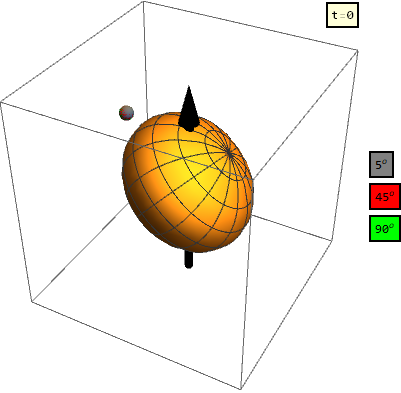}
        \caption{}
    \end{subfigure}
    
    \begin{subfigure}[b]{0.3\textwidth}
        \includegraphics[scale=0.35]{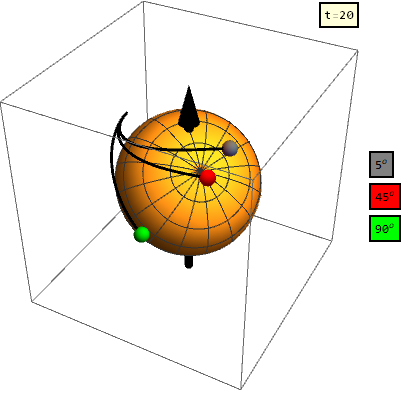}
        \caption{}
    \end{subfigure}
    
        \begin{subfigure}[b]{0.3\textwidth}
        \includegraphics[scale=0.35]{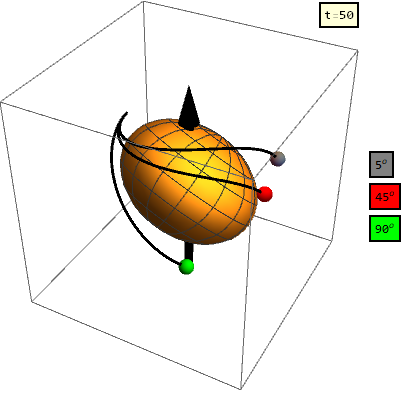}
        \caption{}
    \end{subfigure}
    
%    \centering
%    \includegraphics[width = 2.0in,height=2.0in]{0.png}
    %\hskip .4 cm
%    \includegraphics[width = 2.0in,height=2.0in]{20.png}
    
%    \includegraphics[width = 2.0in,height=2.0in]{50.png}
    
%    \hspace{0.4cm} \scriptsize{(a)} \hspace{0.4cm} \scriptsize{(b)} \hspace{0.4cm} \scriptsize{(c)}

    \caption{(Color online) Snapshots of the simulation of a test particle moving around the misaligned star in actual space (previously, all the calculation and plotting were done in the coordinate system defined around the magnetic field). Here the rotation axis is not perpendicular to the magnetic bulge but at an angle $\chi=45^o$ 
        tilted, however, for comparison, we have also shown the motion of the test particle for $\chi=5^o$ and $\chi=90^o$ misalignment. We have taken magnetic 
        field= $8.0\times10^{17}$ G similar to that of fig \ref{GNL6013}(b). We have taken snapshots for four different instant of time, (a) t=0 sec, (b) t=20 sec, (c) t=50 sec.}
    \label{t0}
\end{figure}

Remember, for simplicity we have solved our equation in magnetic coordinate system.
In fig \ref{t0}, we show the path of three particles in the rotation or the actual coordinate frame, where an arrow shows the rotation axis, and the oblique star is rotating around it. We observe the path of three particles for three different misaligned angles. For $5^{\circ}$ misalignment, the rotation, and magnetic axis are aligned, and for $90^{\circ}$ misalignment, they are perpendicular to each other. We see that the three particles' path differs for different misaligned angles, as discussed in fig \ref{GNL6013}(b). The animation is also provided for this figure in the supplementary file.

From the above discussion, we see that the geodesic of a particle near a rotating NS/magnetar is affected hugely by the GR effect (the FD effect). The rotational drag, along with magnetic distortion of the space-time, bends the particle's path near it. This has significant implications in many astrophysical scenarios starting from electromagnetic emission from the typical pulsar, gamma-ray bursts in magnetars to GW from NSs. In the next two subsections, we discuss the implication of such FD effects in NS physics.

\subsection{Gravitational Wave}
The FD velocity $\beta$ along the polar ($\theta$) direction arising due to the 
oblique rotation of an axisymmetric star significantly modifies the continuous GW coming from a rotating magnetar.
The GW equation can be can be calculated from the change of the time variation of the quadrapole moment of mass distribution given as \cite{wheeler},  $h^{TT}_{ij}=\frac{2G}{c^4}F(i)\frac{1}{r}\frac{d^2I}{dt^2}$, where $F(i)$ is a function of the inclination from observer angle $i$. Now combining equation (\ref{ome}) and (\ref{rot}), we see that the total rotation vector can be written as,
\begin{align}
\vec{\Omega}=&\bar{\alpha}(r)[1+\epsilon]\cos\chi \hat{Z}_M+\bar{\beta}(r) \sin{\chi} \nonumber\\
&(\hat{Y}_M \cos\omega t-\hat{X}_M \sin\omega t)
\label{rott}
\end{align}
where $\epsilon\equiv \left(\frac{I_{zz}-I_{xx}}{I_{0}}\right)$ is the ellipticity and $\bar{\alpha}(r)\equiv \Omega-\alpha(r)$ and  $\bar{\beta}(r)\equiv \Omega-\beta(r)$ are the relative FD terms. Thus the magnitude of the total angular velocity in terms of this relative FD and ellipticity is
\begin{align}
    |\Omega|=\sqrt{\bar{\alpha}^2(1+2\epsilon)\cos^2\chi+\bar{\beta}^2\sin^2\chi}
\end{align} 
Hence the GW amplitude becomes,
\begin{equation}
    h_0=\frac{4G}{c^4}\frac{I\epsilon}{r}\Big[\bar{\alpha}^2(1+2\epsilon)\cos^2\chi+\bar{\beta}^2\sin^2\chi\Big]\sin\chi.
    \label{gw}
\end{equation}  
The relative FD terms could be seen from fig \ref{PLZ2013}, that, as $r\rightarrow\infty$
\begin{align}
\bar{\alpha}(r)\rightarrow\Omega \\
%\end{align}
%\begin{align}
\bar{\beta}(r)\rightarrow\Omega
\end{align}
and the GW amplitude boils down to the standard form \cite{bonazzola,jones,ciolfi}. Usually in the general expression found in the literature regarding the GW amplitude the term involving $\bar{\alpha}$ and $\bar{\beta}$ is not present. However, this expression shows that both the FD velocities play and important role towards determining the strength and nature of GW.
Thus, even for an axisymmetric star, if the rotation axis is not the star's symmetry axis, it loses energy in the form of a GW. The magnitude of the GW depends both on the squares of  $\bar{\alpha}$  and $\bar{\beta}$ terms. However, if $\chi=0$, the rotation axis would be aligned to the symmetry axis (magnetic axis), and thus the star would be axisymmetric with respect to the rotation axis and will not emit any GWs. 
For small misalignment angle $\sin \chi \sim \chi, \cos \chi \sim 1$, the GW amplitude is governed by azimuthal FD velocity but also has small contribuution from polar FD velocity ($h_0=\frac{4G}{c^4}\frac{I\epsilon}{r}[\bar{\alpha}^2(1+2\epsilon)+\bar{\beta}^2\chi^3]$). As the misalignment angle increases the contribution from polar FD velocity increases and completely dominates for a orthogonal rotator.
For the orthogonal rotator the gravitational amplitude has only the polar FD contribution $h_0=\frac{4G}{c^4}\frac{I\epsilon}{r}\bar{\beta}^2$.
The effect of the polar FD term becomes evident when we calculate the electromagnetic energy loss in the next section.

\subsection{Vacuum dipole Magnetosphere}

Recently, calculations have shown that the magnetosphere plays an essential part in determining the pulsars' continuous electromagnetic emission \cite{cerutti1}. Pulsar emission require modeling of 
rotating NS or magnetar and determining the nature of their magnetosphere. 
Calculation has shown that an aligned rotator has zero net poynting flux that is, it doesn't lose any energy \cite{cerutti1,petri}. It was shown that for a misaligned classical oblique rotator, the star does lose its energy \cite{pacini,ostriker} and the power radiated is given by
\begin{align}
    \frac{dE}{dt}=-\frac{2\pi\Omega^4\mu_0}{3c^3}\mu^2\sin^2\chi,
\end{align}
where, $\mu_0$ is the permeability in free space and $\mu$ is the star's magnetic moment.
The classical calculation cannot capture any FD effect which is important in determining the particle distribution in magnetosphere and the energy emitted from the star.
In order to find the loss of energy in a general relativistic case, we first define the magnetic field components in Cartesian components which will be defined by our magnetic potential $a_\phi$ as in equation (\ref{mag3}) and is given by
\begin{align}
    &B^M_x=-\frac{2 a_\phi \cos \theta \sin z}{r^2}\\
    &B^M_y=0\\
    &B^M_z=-\frac{2 a_\phi \cos \theta \cos z}{r^2}
\end{align}
where $z\equiv\left(\frac{a_\phi' \sin \theta e^{-\frac{\lambda}{2}}}{r}\right)$ and $M$ stands for our choice of the magnetic coordinate frame. Converting this magnetic field from our choice of frame of reference $\{X_M,Y_M,Z_M\}$ back into the original reference frame $\{X_R,Y_R,Z_R\}$ , the components of magnetic field is given by 
\begin{align}
    &B_x^R=\frac{2a_\phi \cos\theta \left(\sin\chi \sin\omega t \cos z+\cos \omega t \sin z\right)}{r^2},\\
    &B_y^R=\frac{2 a_\phi \cos\theta \left(\sin\omega t\sin z-\sin\chi \cos \omega t \cos z\right)}{r^2},\\
    &B_z^R=-\frac{2 a_\phi \cos\theta\cos\chi\cos z}{r^2}.
\end{align}

Using equation (\ref{rot}) and (\ref{rott}) to calculate the total induced electric field from the rotating magnetic field, we get
%\begin{widetext}
    \begin{align}
    &E_x^R=\frac{a_\phi\cos\theta \left(-\sin 2\chi \cos \omega t  ((1+\epsilon) \bar{\alpha} (r)+\bar{\beta} (r)) \cos z+2 (1+\epsilon) \bar{\alpha} (r) \cos\chi \sin \omega t \sin z\right)}{r}\\
    &E_y^R=\frac{a_\phi \cos\theta \left(\sin 2 \chi \sin \omega t ((1+\epsilon) \bar{\alpha} (r)-\bar{\beta} (r)) \cos z+2 (1+\epsilon) \bar{\alpha} (r) \cos\chi \cos \omega t \sin z\right)}{r}\\
    &E_z^R=-\frac{2 a_\phi\cos\theta\bar{\beta} (r) \sin\chi \left(\sin \chi \sin 2\omega t\cos z+\cos 2\omega t\sin z\right)}{r}
    \end{align} 
%\end{widetext}

Thus the net poynting flux now can be calculated by integrating it over a sphere of radius $R(\theta)$, which is the radius of the star as a function of $\theta$. Thus the energy loss from the star is given by
%\begin{widetext}
    \begin{align}
    \frac{dE}{dt}=-E_0 \bar{\beta}\left(\frac{(1+\epsilon)}{4}\bar{\alpha}\cos^2\chi \sin2\omega t + 2 \bar{\beta}\sin^2\chi \sin 4\omega t\right)\sin^2\chi
    \end{align}
%\end{widetext}
    where $E_0=-\frac{8 \pi a_\phi^2\omega R}{3\mu_0c^2}$. Hence we see that, as $\chi\rightarrow0$ that is as, the star becomes an aligned rotator, the energy loss $\frac{dE}{dt}\rightarrow0$. To have an overall simple estimation for power loss by a slightly oblique rotator (small $\chi$ limit), we make some simplistic assumption. From eqn (\ref{ome}) we have $\omega = \Omega \epsilon \cos \chi$ and in the small $\chi$ limit it is  $\omega = \Omega \epsilon$. Also for slowly rotating star we have $\Omega$ to be small and if we assume $\epsilon$ also to be small the power loss equation reduces to $\frac{dE}{dt}=-E_0 \bar{\beta} \chi^2 \omega t \left(\bar{\alpha} + 16 \bar{\beta} \chi^2 \right)$ (neglecting O($\epsilon^2$) terms). 
    For an orthogonal rotator the energy loss (in this limit is $\frac{dE}{dt}=-8 E_0 \bar{\beta}^2 \omega t$) is only proprtional to the $\beta$ FD velocity. Therefore, we see that the power loss is determined mostly by the polar FD velocity and the star does not lose energy if $\beta$ goes to zero. 
    This is totally an general relativistic effect which the classical calculation fails to capture.

These two calculation shows that the general relativistic treatment of obliquely rotating magnetar is essential in understanding various observable signatures coming from magnetars. 
We agree that a full 3-dimensional calculation would capture other details; however, our calculation shows that even from an axisymmetric star if we consider a realistic scenario of pulsar lighthouse mechanism, non-axisymmetric FD term generates. This term plays a vital role in modeling observational signatures generated from a rotating NS or magnetar.

\section{Summary and conclusion}

For a highly magnetized slowly rotating magnetar, the magnetic axis becomes the symmetry axis. And if rotational deformation is neglected, we have an axisymmetric star rotating obliquely.
We study such an obliquely rotating star in the GR framework using a perturbative approach. Having the magnetic axis as the symmetry axis, we have decomposed the rotation and the FD arising due to rotation into two-component. One component of the FD is along the azimuthal ($\phi$) direction ($\alpha$) and another in the polar ($\theta$) direction ($\beta$). Solving the Einstein equation perturbatively, we obtain our EOM for different degrees of approximation. We then solve the equations and obtain the geodesic of particles near such an obliquely rotating star. The geodesic of the particle has now contribution from both $\alpha$ and $\beta$ FD.

The two different FD differs because of the star's non-symmetric density distribution (as the star is an oblate spheroid and tilted). We find that $\alpha$ and $\beta$ falls as we go from the center to the star's surface. Due to these two FD effects, the geodesic of a particle depends on the misalignment angle $\chi$ and the magnetic field strength. For $\chi = 0$, we have an aligned axisymmetric rotator, and it does not have the $\beta$ FD, and for $\chi=90^{\circ}$ for the orthogonal rotator there is no $\alpha$ FD. However, to model (light hose effect) pulsar, we have an oblique rotator where both the FD velocities are present. For an oblique rotator, the geodesics of a particle are not planar, and the particles also move along the polar direction.

We have shown that even for an axisymmetric star, the GW is non zero if it rotates obliquely.
For an obliquely rotating axisymmetric star, the continuous GW emitted from the star has its amplitude modulated by the two FD velocities. For an orthogonal rotator, the GW amplitude is proportional to the square of the polar FD velocity. If the misalignment angle is small, the azimuthal FD velocity dictates the GW amplitude. 

The GR calculation shows that the energy loss for an oblique rotator depends both on the $\alpha$ and $\beta$ FD velocity. It goes to zero for an aligned rotator, and for an orthogonal rotator in the small deformation limit, it is only proportional to the $\beta$ FD velocity. The $\beta$ FD velocity dominates the energy loss from an obliquely rotating star, and therefore to ignore this term in the relevant calculation of power loss from a NS could restrict realistic calculation.

We should mention that we have done this calculation with the restriction of the rotational distortions being small and have curtailed our perturbatively expanded term up to second order in a magnetic field and first-order in rotation. However, we have seen that even a 2-dimensional calculation of an obliquely rotating star captures effect in which an axisymmetric star fails. In the immediate future, we would like to implement this calculation to determine the continuous GW from magnetars and extend the vacuum magnetosphere calculation to the force-free magnetosphere and perform a numerical simulation.  

\section*{Acknowledgements}
The authors are grateful to IISER Bhopal for providing all the research and infrastructure facilities. RM would also like to thank the SERB, Govt. of India, for monetary support in the form of Ramanujan Fellowship (SB/S2/RJN-061/2015) and Early Career Research Award (ECR/2016/000161). 
DK also thanks CSIR, Govt. of India for financial support.
%\vspace{-1em}

%%use  somewhere in the left column of the last page to balance the two columns in the end page

%%References section

%\balance

%\begin{widetext}

\section{Appendix}
\appendix

\section{r-Geodesic of order $O(\epsilon_{\Omega}\epsilon_B^2)$}
%\begin{widetext}
	\begin{align}
	r''(t)&=\frac{1}{r^2}e^{-\lambda} \Big[r^2 \Big(\frac{r}{2}\left(3 \cos ^2\theta-1\right) \left(r\frac{dk_2}{dr}+2k_2\right)\left(\theta '(t)^2+\sin ^2\theta \phi '(t)^2\right)-e^{\nu}
	\Big(\frac{dh_0}{dr}+h_0\frac{d\nu}{dr}+\frac{1}{4}(3\cos2\theta+1) \left(\frac{dh_2}{dr}+h_2\frac{d\nu}{dr}\right)\Big)\Big)\nonumber\\
	&+e^{2\lambda} r'(t)\Big[6r\sin
	\theta\cos\theta m_2\theta '(t)-r'(t)\Big(r\frac{dm_0}{dr}+m_0\left(r\frac{d\lambda}{dr}-1\right)+\frac{1}{4}(3\cos2\theta+1)\left(r\frac{dm_2}{dr}+m_2\left(r
	\frac{d\lambda}{dr}-1\right)\right)\Big)\Big]+re^{\lambda}\Big(m_0\nonumber\\
	&+\frac{1}{4}(3\cos2\theta+1) m_2\Big)\Bigg(e^{\nu}\frac{d\nu}{dr}-2r\Big(\theta '(t)^2+\sin ^2\theta
	\phi '(t)^2\Big)\Bigg)\Big]+e^{-\lambda}\Big[\sin\chi\theta '(t) \Bigg(\frac{1}{2} e^{\lambda}\left(r\frac{d\beta_0}{dr}+2\beta_0\right)(4m_0+(3
	\cos2\theta+1)m_2)\nonumber\\
	&-r(2(\frac{1}{4}(3\cos2\theta+1)(r\beta_0\frac{dk_2}{dr}+k_2\left(r\frac{d\beta_0}{dr}+2\beta_0\right))+\beta_1)+r\frac{d\beta_1}{dr})\Bigg)-\frac{1}{2}  \sin ^2\theta\cos\chi\phi '(t) (r (r(3\cos2\theta+1)\nonumber\\ 
	&\alpha_0
	\frac{dk_2}{dr}+(3\cos2\theta+1)k_2\left(r\frac{d\alpha_0}{dr}+2\alpha_0\right)+2r \frac{d\alpha_1}{dr}+4\alpha_1)-e^{\lambda} \left(r\frac{d\alpha_0}{dr}+2\alpha_0\right)(4m_0+(3\cos2\theta+1)m_2))\Big]
	\end{align}
	
%	\end{widetext}

\end{document}